\documentclass[aps,prx,reprint,groupedaddress,amsmath,amssymb,superscriptaddress]{revtex4-2}
\usepackage{graphicx}
\usepackage{float}
\usepackage{multirow}
\usepackage{natbib,twoopt}
\usepackage[breaklinks=true]{hyperref}
\usepackage{xcolor}
\usepackage{amssymb}
\usepackage{graphicx}
\usepackage{amsmath}
\usepackage{hyperref}
\usepackage{color}
\usepackage{subfigure}
\usepackage{verbatim}

\begin{document}

\title{Hybridization of electric and magnetic responses in the effective axion background} 


\author{Timur Z. Seidov}
\email{t.seidov@metalab.ifmo.ru}
\affiliation{School of Physics and Engineering, ITMO University, Saint  Petersburg, Russia}

\author{Maxim A. Gorlach}
\affiliation{School of Physics and Engineering, ITMO University, Saint  Petersburg, Russia}


\begin{abstract}{We investigate the electromagnetic fields produced by the oscillating point electric or magnetic dipole placed in a spherical volume with nonzero time-independent effective axion field. Our analytical solution shows that the fields outside the volume are a superposition of electric and magnetic dipole fields. Such multipole time-dependent generalization of the Witten effect can be realized in magneto-electrics, topological insulators or metamaterials providing a flexible probe of $P$- and $T$-symmetry breaking phenomena in different electromagnetic structures.}
\end{abstract}

\maketitle

\section{Introduction}

Axion is a hypothetical particle originally introduced to solve strong CP problem in quantum chromodynamics~\cite{Peccei_Axion_Origin,Wilczek_Axion_Origins,Weinberg_Axion_Origin}. In parallel, axion has become a prominent candidate for the role of cold dark matter~\cite{Choi_Axion_Review}. Experimental evidence for the existence of axion is currently lacking~\cite{Workman_Holy_Bible}. However, some condensed matter systems provide the realization of the effective axion-like fields. Such effective fields arise in magneto-electrics, multiferroics and  topological insulators~\cite{Rosenberg_Witten_effect,Nenno_Axion_Condensed} as well as in the artificial media~--- metamaterials~\cite{Shaposhnikov_Emergent_Axion,Prudencio_Space-Time_Crystals}. These systems provide a tabletop platform for testing the effects of axion electrodynamics even at optical frequencies \cite{Ahn_Optical_Freq}.

One of the well-celebrated predictions of axion electrodynamics is the emergence of the effective dyon charges induced by real electric charges known as Witten effect \cite{Witten,Wilczek_Applications}. Typically such effect is considered for static charges and static axion fields~\cite{Rosenberg_Witten_effect} though generalization towards time-varying axion fields has recently been done \cite{Hill_Dipole_Moment} with the proposal to use such an effect in cosmic axions detection.

While both theoretical and experimental studies indicate that the effective axion response of the material for static electric field is indeed possible~\cite{Uri2020-Static_case}, a significant part of condensed matter systems and artificial materials feature vanishing effective axion field at zero frequencies. Therefore, the response of such materials to time-varying sources and currents needs to be carefully examined. To fill this gap, here we investigate the fields produced by the oscillating electric or magnetic dipole surrounded by the spherical volume featuring effective axion response. We identify the signatures of the effective axion response in this case showing that our treatment can be readily generalized towards the arbitrary multipole source providing rather general, yet simple insight into the underlying physics.

The rest of the article is organized as follows: in Section~\ref{sec:Problem_formulation} we provide the general formulation of the problem. In the next Section~\ref{sec:Static_Case} we examine the static limit of electromagnetic fields in our system to analyze the role of the boundary conditions and obtain the results for a simple limiting case. In the following Section~\ref{sec:Nonstatic_case} we analyze the case of oscillating dipoles employing the multipole expansion technique. We show that in case of oscillating dipoles, the field outside of the axion surroundings is presented as a superposition of electric and magnetic dipole fields with orthogonal polarizations and distinct frequency dependencies. In Section~\ref{sec:Eigenmodes} we consider the eigenmodes of the axion shell and their link to the obtained solution. We conclude our work with the Section~\ref{sec:Discussion} discussing the results and their possible generalizations and consequences.

\section{Problem formulation}~\label{sec:Problem_formulation}

It is known that the equations of axion electrodynamics can be recast as  Maxwell's equations in the medium with special constitutive relations~\cite{Nenno_Axion_Condensed}:
\begin{gather}
    \mathbf{D}=\varepsilon \mathbf{E} + \chi \mathbf{B}, \label{ConstRel1} \\ 
    \mathbf{H}=-\chi \mathbf{E} + \mu^{-1} \mathbf{B}. \label{ConstRel2}
\end{gather}
Here, $\chi$ is a real parameter quantifying the strength of the effective axion response often referred to as Tellegen or non-reciprocity parameter. Typical values of $\chi$ remain well under unity \cite{Pyatakov2012,Nenno_Axion_Condensed}, though there are no fundamental restrictions on its magnitude. 

As can be readily verified, time-independent homogeneous $\chi$ is not manifested in the bulk but yields nontrivial effects at the boundary, since Maxwell's equations involve only spatial and temporal derivatives of $\chi$. Boundary conditions at the interface with vacuum read
\begin{align}
    [\mathbf{B}]_n=0, & \quad & [\mu^{-1}\,\mathbf{B}]_t=\chi \mathbf{E}_t, \label{AxionBC1} \\
    [\mathbf{E}]_t=0, & \quad & [\varepsilon\,\mathbf{E}]_n=-\chi \mathbf{B}_n, \label{AxionBC2}
\end{align}
where the notation $[\mathbf{A}]_{n,t}$ stands for the difference of normal or tangential components in the structure and vacuum, respectively.

\begin{figure}
    \centering
    \includegraphics[scale=0.2]{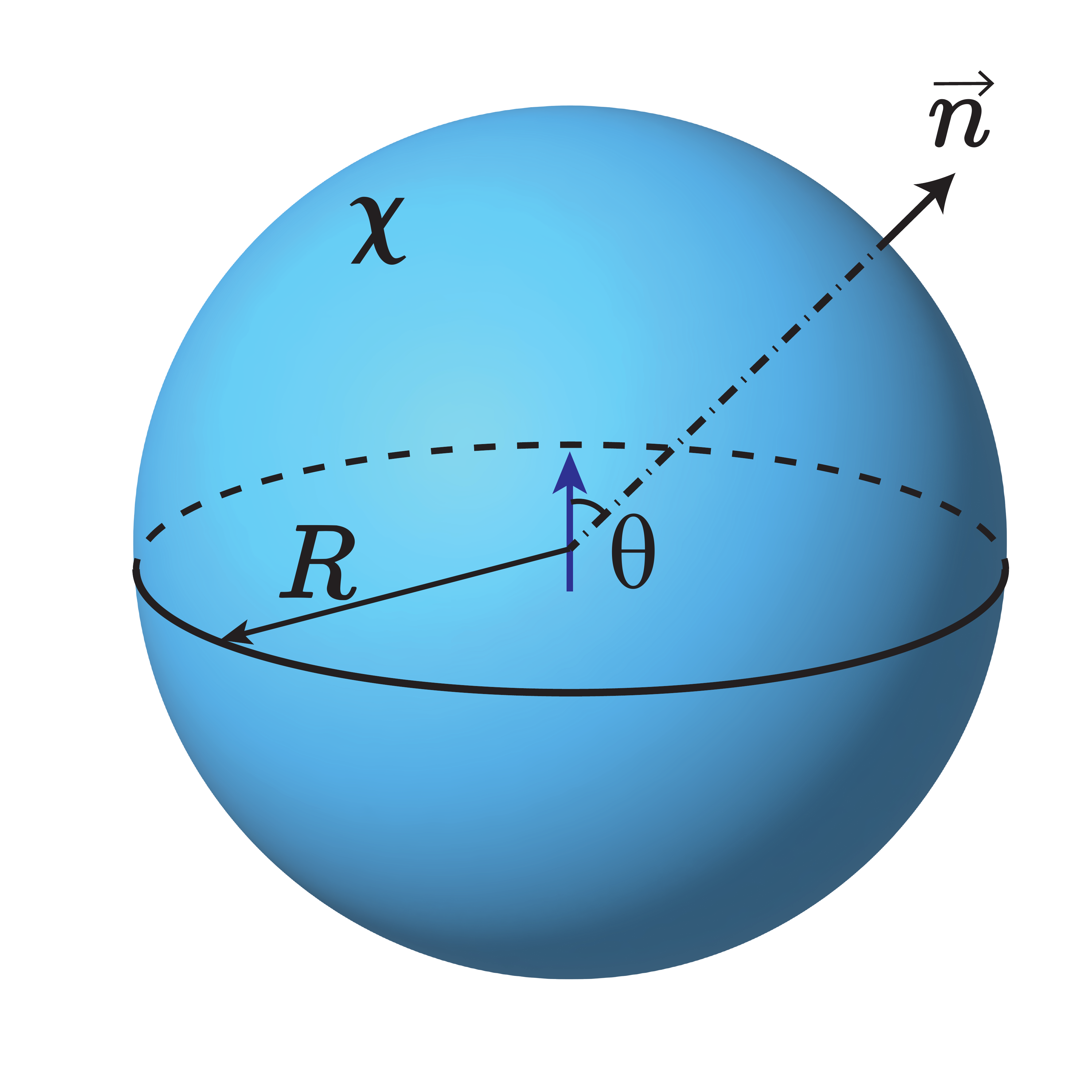}
    \caption{Geometry of the problem. Electric or magnetic dipole moment, depicted as blue arrow, resides inside a spherical volume with the radius $R$ featuring effective axion response $\chi$.}
    \label{ProblemGeom}
\end{figure}

To examine the role of the effective axion background, we analyze the following problem (Fig.~\ref{ProblemGeom}). A point dipole is located at the origin oscillating with the frequency $\omega$. It is surrounded by the medium with Tellegen response $\chi$. Below, we calculate the electromagnetic field produced by this dipole both inside and outside the axion medium. To simplify the treatment, we assume that the material has a spherical shape with the radius $R$, which allows us to treat the problem analytically providing the insight into physics at play.

\section{Solution for the static case}~\label{sec:Static_Case}

First, to provide a simple and intuitive picture, we analyze the static limit $\omega\rightarrow 0$ assuming for simplicity $\varepsilon=\mu=1$ for the axion material. In such case, the field of the static dipole, e.g. electric one, is defined as
\begin{equation}
    \mathbf{E}_{\text{in}}=\frac{1}{r^3}[3(\mathbf{n}\cdot\mathbf{d}_0)\mathbf{n}-\mathbf{d}_0]+\mathbf{E}_0.
\end{equation}
The first term here is the field of point electric dipole with a dipole moment $\mathbf{d}_0$, $\mathbf{n}$ being a unit vector collinear to the radius vector. The second term describes the field induced by the bound charges at the spherical interface with vacuum. In addition, we expect that bound currents at the interface will induce the magnetic field inside $\mathbf{B}_{\text{in}}=\mathbf{B}_0$. 

In turn, we seek the field outside the axion material as a superposition of both electric and magnetic dipole fields:
\begin{gather}
    \mathbf{E}_{\text{out}}=\frac{1}{r^3}[3(\mathbf{n}\cdot\mathbf{d})\mathbf{n}-\mathbf{d}] , \\
    \mathbf{B}_{\text{out}}=\frac{1}{r^3} [3(\mathbf{n}\cdot\mathbf{m})\mathbf{n}-\mathbf{m}] .
\end{gather}
Thus, we have four equations given by boundary conditions \eqref{AxionBC1},\eqref{AxionBC2} and four unknowns to determine: $\mathbf{E}_0,\mathbf{B}_0,\mathbf{d}$ and $\mathbf{m}$. Substituting the ansatz for the fields into the boundary conditions, we recover
\begin{gather}
    \mathbf{n}\cdot \mathbf{B}_0=\frac{2(\mathbf{m}\cdot \mathbf{n})}{R^3}, \\
    -\frac{\mathbf{n}\times\mathbf{d}_0}{R^3}+\mathbf{n}\times\mathbf{E}_0=-\frac{\mathbf{n}\times\mathbf{d}}{R^3}, \\
    \mathbf{n}\times\mathbf{B}_0-\chi\left[\mathbf{n}\times\mathbf{E}_0-\frac{\mathbf{n}\times\mathbf{d}_0}{R^3}\right]=-\frac{\mathbf{n}\times\mathbf{m}}{R^3}, \\
    \frac{2(\mathbf{n}\cdot\mathbf{d}_0)}{R^3}+\mathbf{n}\cdot\mathbf{E}_0+\chi\mathbf{n}\cdot\mathbf{B}_0= \frac{2(\mathbf{n}\cdot\mathbf{d})}{R^3}.
\end{gather}
Since the boundary is a closed surface, $\mathbf{n}$ is an arbitrary unit vector, so multiplication by it can be omitted. Straightforward calculation yields 
\begin{align}
    \mathbf{d}=\frac{\mathbf{d}_0}{1+\frac{2\chi^2}{9}}, & \quad & \mathbf{m}=-\frac{\chi}{3}\frac{\mathbf{d}_0}{1+\frac{2\chi^2}{9}}, \label{MD_zero_freq} \\
    \mathbf{E}_0=\frac{2\chi^2}{9 R^3}\frac{\mathbf{d}_0}{1+\frac{2\chi^2}{9}}, & \quad & \mathbf{B}_0=-\frac{2\chi}{3 R^3}\frac{\mathbf{d}_0}{1+\frac{2\chi^2}{9}}.
\end{align}

A similar calculation for the magnetic moment $\mathbf{m}_0$ inside the spherical axion material yields the expressions
\begin{align}
    \mathbf{d}=\frac{2\chi}{3}\frac{\mathbf{m}_0}{1+\frac{2\chi^2}{9}}, & \quad & \mathbf{m}=\frac{\mathbf{m}_0}{1+\frac{2\chi^2}{9}}, \\
    \mathbf{E}_0=-\frac{2\chi}{3 R^3}\frac{\mathbf{m}_0}{1+\frac{2\chi^2}{9}}, & \quad & \mathbf{B}_0=-\frac{4\chi^2}{9 R^3}\frac{\mathbf{m}_0}{1+\frac{2\chi^2}{9}}.
\end{align}

In the following we will refer to moments ${\bf d}_0$ and ${\bf m}_0$ inside the material as seed moments and to $\mathbf{d},\mathbf{m}$ moments as induced moments.

Thus, we observe that the static axion surroundings of magnetic dipole induce an effective electric dipole moment perceived by the outside observer. Note that the similar result has been obtained for electron in the oscillating axion field~\cite{Hill_Dipole_Moment}. In the same way, static axion material surrounding the electric dipole induces an effective magnetic dipole moment.


\section{Solution for the time-varying dipole moments}~\label{sec:Nonstatic_case}

Given the symmetry of the problem and the obtained solution for the static limit, we seek the general solution in the form of multipole expansion utilizing the notion of vector spherical harmonics (VSH)~\cite{jackson}:
\begin{gather}
        \mathbf{E}=\sum_{l,m}\left\lbrace\frac{i}{q}\nabla\times[a_E(l,m) N_l\mathbf{X}_{l,m}] - a_M(l,m) M_l \mathbf{X}_{l,m}\right\rbrace, \\
        \mathbf{B}=\sum_{l,m}\left\lbrace a_E(l,m) N_l\mathbf{X}_{l,m} + \frac{i}{q}\nabla\times[a_M(l,m) M_l \mathbf{X}_{l,m}]\right\rbrace,
\end{gather}
where $q=\omega/c$, $N_l,M_l$ are the linear combinations of spherical Bessel and Hankel functions of the $l$-th order, $\mathbf{X}_{l,m}(\theta,\varphi)$ is a normalized VSH defined as follows
\begin{equation}
    \mathbf{X}_{l,m}(\theta,\varphi)=\frac{1}{\sqrt{l(l+1)}}\hat{\mathbf{L}} Y_{l,m}(\theta,\varphi).
\end{equation}
Here, $Y_{l,m}(\theta,\varphi)$ is a spherical harmonic and $\hat{\mathbf{L}}=-i{\bf r}\times\nabla$ is an angular momentum operator. Such normalized VSH are orthonormal with respect to $L^2$ functional measure and possess a useful property: for any $f(qr)$
\begin{multline}
    \nabla\times[f(qr)\mathbf{X}_{l,m}]=\frac{1}{r}\underbrace{\frac{d}{dr}[r f(qr)]}_{D_r f(qr)} \underbrace{\mathbf{n}\times\mathbf{X}_{l,m}}_{\mathbf{Z}_{l,m}}+ \\ i\sqrt{l(l+1)}\frac{f(qr)}{r}Y_{l,m}\mathbf{n}. \label{VSHprop}
\end{multline}
Note that $\mathbf{n}$ is a unit vector collinear with the radius vector which, as before, coincides with the normal to the spherical boundary. For brevity, we also introduce special derivative operator $D_r$ and $\mathbf{Z}_{l,m}$ VSH. By definition, vector harmonics $\mathbf{Z}_{l,m}$ are also orthonormal and, together with $\mathbf{X}_{l,m}$, form a basis in tangent to spherical boundary space for each particular set of indices $l,m$. Each VSH of the $l$-th order corresponds to the multipole of the same order, while $m$ index spans the range from $-l$ to $l$.

In the case of dipole, $l=1$ and dipole moment components are related to the multipole coefficients as $a_E(1,1)=iq^3\sqrt{4\pi/3}\,(d_x-id_y), \  a_E(1,0)=-iq^3\sqrt{8\pi/3}\,d_z, \ a_E(1,-1)=a_E(1,1)^*$. Note that the dipole moment of the system is defined here based on its far field structure which is in line with the works~\cite{Alaee_Exact_Multipoles,Fernandez-Corbaton_Exact_Multipoles}.

Since the medium has full rotational symmetry and $\mathbf{X}_{l,m},\mathbf{Z}_{l,m}$ are orthonormal, the boundary conditions for each set of $(l,m)$ indices are independent.

As radial functions $N,M$ inside the axion medium, we use the following ones: $N_l=(A h^{(1)}_l(qr)+B j_l(qr))$ and $M_l=\tilde{B} j_l(qr)$. Here,   $A,B,\tilde B$ are the constants to determine, and $j_l, h^{(1)}_l$ are spherical Bessel function and spherical Hankel function of the first kind, respectively. Such choice of the radial functions is due to the fact that $h_l^{(1)}$ corresponds to the outgoing wave radiated by the multipole of the order of $l$, while $j_l$ is regular at zero and hence can describe the boundary-induced field. Thus, if $z$ axis is chosen parallel to the dipole, our anzatz for the solution inside reads
\begin{widetext}
    \begin{gather}
        \mathbf{E}_I=\frac{i}{q}\nabla\times[(A h_1^{(1)}(qr) + B j_1(qr))\mathbf{X}_{10}(\theta,\varphi)]- 
        \tilde B j_1(qr) \mathbf{X}_{10}(\theta,\varphi), \label{E_expansion} \\
        \mathbf{B}_I=(A h_1^{(1)}(qr) + B j_1(qr))\mathbf{X}_{10}(\theta,\varphi) +   \frac{i}{q}\nabla\times[\tilde B j_1(qr) \mathbf{X}_{10}(\theta,\varphi)]. \label{B_expansion}
    \end{gather}
\end{widetext}
The solution outside of the material is described in terms of outgoing spherical waves:
    \begin{gather}
        \mathbf{E}_O=\frac{i}{q}\nabla\times[a_E h_1^{(1)}(qr)\mathbf{X}_{10}(\theta,\varphi)] -  a_M h_1^{(1)}(qr) \mathbf{X}_{10}(\theta,\varphi), \\
        \mathbf{B}_O=a_E h_1^{(1)}(qr)\mathbf{X}_{10}(\theta,\varphi) + \frac{i}{q}\nabla\times[ a_M h_1^{(1)}(qr) \mathbf{X}_{10}(\theta,\varphi)]. 
    \end{gather}
For the sake of brevity, from now on we omit the indices $l,m$ (which are equal to $1,0$) as well as the arguments of VSH and radial functions.

Since the fields are monochromatic, it is sufficient to satisfy the boundary conditions only for the tangential components of the fields, while the conditions for the normal components are satisfied automatically. Furthermore, since $\mathbf{X},\mathbf{Z}$ form a basis in the tangential space, we need to separate $\mathbf{X}$- and $\mathbf{Z}$-components, using Eq.~\eqref{VSHprop}. Combining the result with boundary conditions Eqs.~\eqref{AxionBC1},\eqref{AxionBC2}, we recover the system of equations
    \begin{gather}
        a_M=\tilde B \frac{j}{h^{(1)}}, \\
        a_E=A+B\frac{D_R j}{D_R h^{(1)}}, \\
        (A h^{(1)} + B j)- a_E h^{(1)}= -\chi \tilde B j, \\
        \tilde B D_R j - a_M D_R h^{(1)}= \chi (A D_R h^{(1)} + B D_R j).
    \end{gather}
Straightforward calculation yields the solution:
    \begin{gather}
        a_M=-\frac{\chi C_1(C_2+1)}{C_1 C_2 + \chi^2} A, \label{M_finite_freq} \\
        a_E=\left(1 - \frac{\chi^2}{C_1 C_2 + \chi^2} \right) A, \label{D_finite_freq} \\
        B= \frac{\chi C_2}{C_1 C_2 + \chi^2}\frac{D_R h^{(1)}}{D_R j} A,\\
        \tilde B= - \frac{\chi^2}{C_1 C_2 + \chi^2}\frac{D_R h^{(1)}}{D_R j} A,
    \end{gather}
where $A$ is proportional to the seed electric dipole moment $\mathbf{d}_0$ as discussed above, and $C_1, C_2$ are defined as 
    \begin{gather}
         C_1=\frac{D_R j}{D_R h^{(1)}}\frac{h^{(1)}}{j}-1, \\
         C_2=\frac{D_R h^{(1)}}{D_R j}\frac{j}{h^{(1)}}-1.
    \end{gather}

The results of the previous section can be readily recovered by expanding the coefficients $C_{1,2}$ in the vicinity of $qR=0$, which yields $C_1 \rightarrow -3$ and $C_2  \rightarrow -3/2$. Substituting these approximate expressions into Eqs.~\eqref{M_finite_freq}, \eqref{D_finite_freq}, we recover Eq.~\eqref{MD_zero_freq}. 

Since the multipole moments are directly related to the coefficients of the multipole expansion, we can readily determine the ratios between seed dipole moment and induced dipole moments as

\begin{gather}
    \frac{\mu_{\text{ind}}}{d_{\text{seed}}}=\frac{a_M}{A}=-\frac{\chi C_1(C_2+1)}{C_1 C_2 + \chi^2}, \\
    \frac{d_{\text{ind}}}{d_{\text{seed}}}=\frac{a_E}{A}=\left(1 - \frac{\chi^2}{C_1 C_2 + \chi^2} \right).
\end{gather}

\begin{figure}
    \centering
    \includegraphics[scale=0.45]{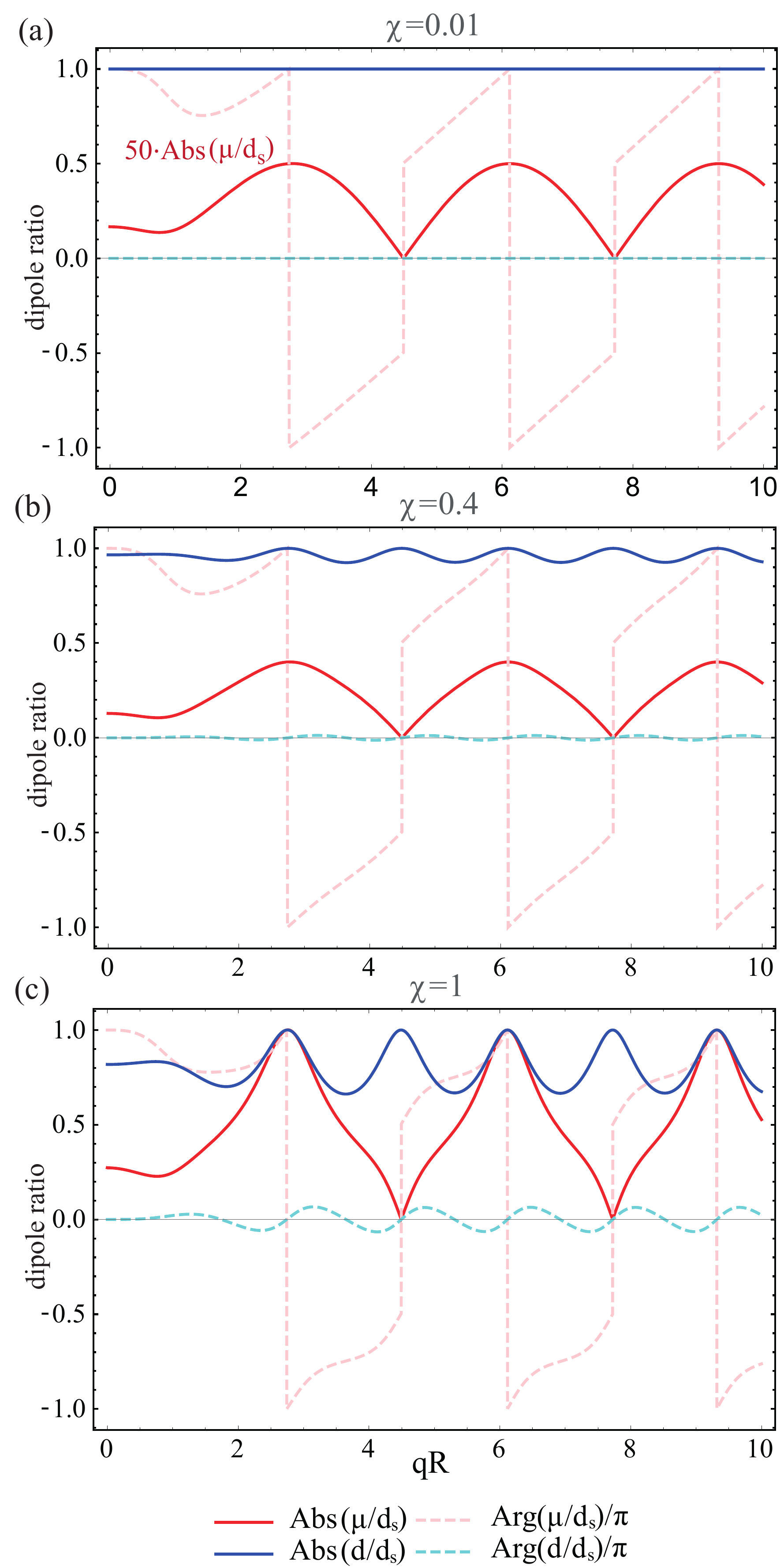}
    \caption{The ratios of the induced electric and magnetic dipole moments $\mathbf{d}$ and $\mathbf{m}$ to the seed electric dipole moment $\mathbf{d}_s$ versus the dimensionless frequency $qR$ calculated for the different values of $\chi$. For $\chi<1$, the induced magnetic moment is always smaller than the induced electric moment (a),(b). For $\chi=0.01$ $\mu_{\text{ind}}/d_{\text{seed}}$ ratio is additionally multiplied by the factor of $50$ for better visibility (a).}
    \label{ratios}
\end{figure}

The calculated ratios are plotted in Fig.~\ref{ratios}. Depending on the chosen frequency, we observe a pattern of maxima and minima for the induced magnetic magnetic moment, and the positions of such specific points are nontrivial, i.e. they do not coincide with $qR=2\pi n$. 

Second, we observe that for $\chi<1$ the induced magnetic moment is below the induced electric moment and does not exceed the seed electric moment at any frequency. What is less intuitive, is that for $\chi>1$ the induced magnetic moment may exceed the induced electric one in certain ranges of frequencies, and the higher is the value of $\chi$, the wider are these ranges. 

Finally, the zeros of the induced magnetic dipole moment can be interpreted as {\it anapole states} of the system~\cite{Mirosh2015,Baryshnikova2019,Yang2019,Savinov2019-Anapoles}. Clearly, the currents are present at the spherical boundary, as dictated by the boundary conditions, but they do not radiate to the far field at these particular frequencies thus creating a non-radiating configuration. Such zeros in magnetic dipole radiation, in fact, coincide with the zeros of $j_1(qr)$ function which in turn guarantee vanishing induced magnetic dipole moment.


\section{Eigenmodes of axion sphere}~\label{sec:Eigenmodes}

Using the same procedure, we may calculate the dipole eigenmodes of the axion sphere in the absence of any external sources. To that end, we need to use only spherical Bessel functions $j$ in the anzatz for the field inside the sphere, while keeping the outgoing spherical waves $h^{(1)}$ for the solution outside. This way we arrive to the system 

\begin{gather}
    \tilde a_M=b_M \frac{j}{h^{(1)}}, \label{EgM1}\\
    \tilde a_E=b_E \frac{D_R j}{D_R h^{(1)}}, \label{EgM2}\\
    b_E j-a_E h^{(1)}=-b_M j\,\chi \label{EgM3}\\
    b_M D_R j- a_M D_R h^{(1)}=b_E D_R j\, \chi ,  \label{EgM4}
\end{gather}
where $b_E$ and $b_M$ quantify the amplitudes of electric and magnetic modes, respectively. Combining the equations, we recover the condition for the eigenmodes 
\begin{gather}
    C_1 C_2 +\chi^2=0,
\end{gather}
and express the coefficients $b_M,a_E,a_M$ via $b_E$, which is arbitrary.
\begin{gather}
    b_M=b_E\frac{\chi}{C_2}, \\
    \tilde a_E=b_E\frac{D_R j^{(1)}}{D_R h^{(1)}}, \\
    \tilde a_M=b_E\frac{\chi}{C_2}\frac{j^{(1)}}{h^{(1)}}.
\end{gather}

In this setting, there are no dipoles inside, but the ratio of the effective induced dipole moments $\tilde\mu, \tilde d$ outside of the sphere is well-defined:
\begin{equation}
    \frac{\tilde \mu_{\text{ind}}}{\tilde d_{\text{ind}}}=\frac{\tilde a_M}{\tilde a_E}=\frac{a_M}{a_E}=\frac{\mu_{\text{ind}}}{d_{\text{ind}}}=-\chi \frac{C_2+1}{C_2}.
\end{equation}

This analysis suggests that the observed behavior of the dipole moments is associated with the eigenmode structure being the property of axion material and its geometry.


\section{Discussion and conclusions}\label{sec:Discussion}

In conclusion, we have analyzed the interplay between time-varying sources such as electric and magnetic dipoles with the static effective axion fields which arise in magneto-electrics, multiferroics, topological insulators or metamaterials. As we prove, this generalization of the well-celebrated Witten effect features the nontrivial dependence on frequency with the induced magnetic moment vanishing at certain frequencies corresponding to the magnetic anapole.

Our results are readily generalized towards higher-order multipoles by simply changing the indices $(l,m)$. This becomes possible due to the fact that the boundary of the axion material does not mix the modes with different $(l,m)$. Note also that the ratios of the quadrupole moment components are independent on $m$.

Moreover, our results can be also generalized to other geometries such as cylindrical one. In such case, vector spherical harmonics need to be replaced by their cylindrical counterparts. Another interesting research direction is to try to find a mapping between solutions with static multipoles and dynamic axion fields and the solutions with static axion fields and time-varying multipole moments.

From the practical perspective, the discussed frequency dependence of Witten effect might be useful: even vanishing Witten effect at certain frequency still does not mean that the effective axion-like fields in the medium are absent.

In general, the obtained results provide a theoretical background to experimentally explore the effects of axion electrodynamics in artificial media or to access the effective axion properties of the material. Nonreciprocal metastructures featuring effective axion responce may also provide a novel twist in Mie-resonant photonics~\cite{Mie_optics}.

\section*{Acknowledgments}

This work was supported by Priority 2030 Federal Academic Leadership Program. The authors acknowledge partial support by the Foundation for the Advancement of Theoretical Physics and Mathematics ``Basis''.

\bibliography{bibliography}

\end{document}